\documentclass[conference]{IEEEtran}
\IEEEoverridecommandlockouts
\usepackage{amsmath}
\usepackage{graphicx}
\usepackage{amssymb}
\usepackage{multicol}
\usepackage{mathrsfs}
\usepackage{multirow}
\usepackage{booktabs}
\usepackage[T1]{fontenc}
\usepackage{caption}
\usepackage{hyperref}
\usepackage{subfigure}
\usepackage{float}
\usepackage{enumerate}
\usepackage{nomencl}
\usepackage{ragged2e}
\makenomenclature

\makeatletter
\newcommand{\linebreakand}{%
  \end{@IEEEauthorhalign}
  \hfill\mbox{}\par\vspace{-3mm}
  \mbox{}\hfill\begin{@IEEEauthorhalign}
}

\DeclareCaptionLabelFormat{smallcaps}{\textsc{#1}~\textsc{#2}}
\DeclareMathSizes{10}{10}{7}{5}  
\newcommand{\diagonal}{\mathrel{\rotatebox[origin=c]{45}{$\leftrightarrow$}}}
\newcommand{\antidiagonal}{\mathrel{\rotatebox[origin=c]{135}{$\leftrightarrow$}}}

\begin{document}
\title{Experimental free-space quantum key distribution over a turbulent high-loss channel\\
\thanks{This material is based upon work supported by the  Department of Energy, Office of Science, Office of Advanced Scientific Computing Research, through the Quantum Internet to Accelerate Scientific Discovery Program under Field Work Proposal 3ERKJ381. We acknowledge support by the National Science Foundation under grant DGE-2152168.}
}
     
\author{\IEEEauthorblockN{Md Mehdi Hassan}
\IEEEauthorblockA{\textit{Department of Physics and Astronomy} \\
\textit{The University of Tennessee}\\
Knoxville, TN, USA \\
mhassa11@vols.utk.edu}\\
\and
\IEEEauthorblockN{Kazi Reaz}
\IEEEauthorblockA{\textit{Department of Physics and Astronomy} \\
\textit{The University of Tennessee}\\
Knoxville, TN, USA \\
kreaz@vols.utk.edu}
\and
\IEEEauthorblockN{Adrien Green}
\IEEEauthorblockA{\textit{Department of Physics and Astronomy} \\
\textit{The University of Tennessee}\\
Knoxville, TN, USA \\
agreen91@vols.utk.edu}

\linebreakand  
\IEEEauthorblockN{Noah Crum}
\IEEEauthorblockA{\textit{Department of Physics and Astronomy} \\
\textit{The University of Tennessee}\\
Knoxville, TN, USA \\
ncrum@vols.utk.edu}
\and
\IEEEauthorblockN{George Siopsis}
\IEEEauthorblockA{\textit{Department of Physics and Astronomy} \\
\textit{The University of Tennessee}\\
Knoxville, TN, USA \\
siopsis@tennessee.edu}
}

\maketitle

\begin{abstract}
Free-space quantum cryptography plays an integral role in realizing a global-scale quantum internet system. Compared to fiber-based communication networks, free-space networks experience significantly less decoherence and photon loss due to the absence of birefringent effects in the atmosphere. However, the atmospheric turbulence contributes to deviation in transmittance distribution, which introduces noise and channel loss. Several methods have been proposed to overcome the low signal-to-noise ratio. Active research is currently focused on establishing secure and practical quantum communication in a high-loss channel, and enhancing the secure key rate by implementing bit rejection strategies when the channel transmittance drops below a certain threshold. By simulating the atmospheric turbulence using an acousto-optical-modulator (AOM) and implementing the prefixed-threshold real-time selection (P-RTS) method, our group performed finite-size decoy-state Bennett-Brassard 1984 (BB84) quantum key distribution (QKD) protocol for 19 dB channel loss. With better optical calibration and efficient superconducting nano-wire single photon detector (SNSPD), we have extended our previous work to 40 dB channel loss characterizing the transmittance distribution of our system under upper moderate turbulence conditions.  \\
\end{abstract}

\hrule
\vspace{1 mm}

\section{Introduction}

Over the course of a year, the number of qubits in the most powerful quantum computers has tripled from 127 to 433.  As engineering and technological capabilities keep increasing, the exponential growth of the number of qubits in quantum computers is inevitable. The relevance of switching from traditional modern encryption to quantum computer-proof data encryption is evident.  

Quantum key distribution (QKD) offers absolute security for distribution encryption keys between two parties against eavesdropping by relying on the principles of quantum mechanics. Since the proposal of the very first quantum key distribution by Bennett and Brassard in 1984 (BB84) \cite{bib:1}, numerous improvements have been made to this kernel, enabling its transition from theory to practical use. The first successful experimental prototype of BB84 \cite{bib:2} was announced in 1989, affirming 32 cm free-space data transmission that exploited the polarization property of light particles (photons). The experiment was done using incoherent green light (550 nm) as the source and photo multiplier tubes as the detectors.

In \cite{bib:1}, the authors addressed the technical difficulties of producing consistent light pulses containing single photons, and proposed coherent or incoherent sources for light pulses. Processes to create single photons, such as through spontaneous parametric down conversion \cite{bib:3} or quantum dots \cite{bib:4}, are expensive and still under active research. Meanwhile, coherent state implementations generate signals at a high rate and utilize cheaper off-the-shelf equipment \cite{bib:5}. Because of its practicality and high secure key rates, BB84 with weak coherent states has been the subject of extensive research and development. An important concern has been the fact that coherent states may contain multiple photons, which an eavesdropper can remove and copy through a Photon Number Splitting (PNS) attack \cite{bib:20}. To thwart such attacks, the decoy state method was invented \cite{bib:6} allowing communicating parties to monitor whether or not Eve has changed the photon number statistics of the light pulses. Over the years, security proofs have become more sophisticated in order to account for an increasing amount of details related to real implementations, such as the finite size effect \cite{bib:21} and the consideration of various side channel and individual attacks \cite{bib:7}. Today, QKD protocols encoded on weak coherent states have well established security proofs, high data rates and have been implemented in a number of environments \cite{bib:25,bib:26,bib:27,bib:28}. While fiber based networks have seen significant improvements over the last few decades, however, transmission losses in standard fiber scale as $ \sim \!\! 0.2 $ dB/km at the telecom wavelength ($\sim \!\!1550$ nm) \cite{bib:8}, and thus the systems are only viable at the intra-metropolitan scale \cite{bib:7}. Therefore, for longer distance secure communication such as a global scale quantum network, free space quantum channels are essential.

Some of the main challenges of free space communication are due to the effects of atmospheric turbulence. Even though the signal can propagate almost without attenuation in the upper atmosphere, the lower levels, especially the troposphere and stratosphere, degrade the intensity of the signal. For example, at an altitude of 1,200 km, the estimated loss due to atmospheric absorption and turbulence ranges between 3 and 8 dB \cite{bib:23,bib:23a}. The variation in air temperature and pressure gradients with respect to position results in eddy flows of air, which cause changes in the refractive index over time and across different locations \cite{bib:9}. This results in beam front deformation and beam wandering, which affects the transmittance.

In this experiment, we simulate finite-key, polarization encoded BB84 with weak laser pulses with cutting edge detectors in a turbulent, high loss, free-space environment. Since atmospheric turbulence can be measured with a classical laser probe, resulting transmittance information can be used to reject groups of coherent states that are most likely to have suffered high loss. Detections that occurred during those time intervals are more likely to be detector noise than the signal and can be rejected to lower the average quantum bit error rate. In \cite{bib:17}, it was argued that the cutoff is dependent only on device parameters rather than channel statistics, meaning that an optimal threshold can be prefixed, reducing data storage and computational requirements, which we investigate experimentally.


\section{Theory}

In this section, we discuss the relationship between a turbulent atmosphere with finite key decoy state parameters for BB84, and the resulting effects to the secure key rate.

\subsection{Atmospheric Effect}
Atmospheric turbulence causes random transmittance fluctuations that can be modeled by a log-normal probability distribution \cite{bib:14} whose density function is given by 
\begin{align}
 p_{_{\eta_o, \sigma}}(\eta)(\eta) = \dfrac{1}{\sqrt{2\pi} \sigma \eta } \exp{{ \left\{ -\frac{[\ln{(\frac{\eta}{\eta_o})}+ \frac{\sigma^2}{2}]^2}{2\sigma^2} \right\} } } \ ,
 \label{eq:1}
\end{align}
where $\eta_o$ is the average channel transmittance and the variance, $\sigma^2$, gives the level of turbulence. $\sigma^2$ is commonly referred to as the Rytov parameter. If the transmitter and the receiver are separated by a horizontal distance $L$, the Rytov parameter is given approximately by the expression $\sigma^2 = 1.23\ C_n^2  k^{\frac{7}{6}} L^{\frac{11}{6}}$ for plane waves with wave number $k$ \cite{bib:12}. Here, $C_n^2$ is the refractive index structure parameter and $n$ is the refractive index of the medium. The structure parameter varies depending on altitude and temperature as well \cite{bib:16}. For example, $\sigma^2=0.924$ corresponds to both a small turbulence of $C_n^2 = 10^{-17}\ \text{m}^{-\frac{2}{3}}$ covering air to air 100 km distance \cite{bib:15} as well as a higher turbulence of $C_n^2 = 6.2\times10^{-15}\ \text{m}^{-\frac{2}{3}}$ covering a distance of $3$ km.

In our experiment, a fiber-coupled acousto-optic modulator (AOM) was used to vary the photon intensity according to our desired transmittance model. We emulated the log-normal distribution \eqref{eq:1} of the output signal via controlling the AOM device by an arbitrary waveform generator (AWG3). The turbulence parameter $\sigma^2$ was set to 1, which corresponds to upper-medium turbulence. 

\begin{table}[ht!]
\caption*{\centering  \textbf{Nomenclature}}
\begin{tabular}{r l} 
\hline \hline \addlinespace[1ex]  
$N$ & Total pulse sent to Bob\\
$R$ & Key rate \\ 
$\ell$ & Number of distilled secure bits\\ 
$\phi_X$ & Phase error \\
$e_{\text{obs}}$ & Observed error \\
$\eta$ & Channel transmittance  \\
$\eta_d$ & Detector efficiency \\
$\mu_i $ & Average photon number per pulse   \\
$q_x$ & Alice's probability of choosing $\boxplus$ basis   \\
$q_z$ & Alice's probability of choosing $\boxtimes$ basis   \\
$P_{\mu_i}$ & Alice's probability of choosing intensity $\mu_i$  \\
$nX_{\mu_i}$ & Detections when both chose basis $\boxplus$ and intensity $\mu_i$  \\
$nZ_{\mu_i}$ & Detections when both chose basis $\boxtimes$ and intensity $\mu_i$  \\
$mX_{\mu_i}$ & Detections in error for basis $\boxplus$ and intensity $\mu_i$  \\
$mZ_{\mu_i}$ & Detections in error for basis $\boxtimes$ and intensity $\mu_i$  \\\addlinespace[1ex]  

\hline \hline
\end{tabular}
\end{table}

\subsection{Key Rate}

As we move from the asymptotic case \cite{bib:10} to a real-life scenario with finite key size, the key rate $R$ becomes dependent on the size of the key. According to the modification proposed in \cite{bib:17}, adopted from \cite{bib:18}:
\begin{flalign}
 R_{\text{GLLP}}\big( \eta \big) \xrightarrow{} R_{\text{fin}}\big(\eta_{\text{avg}}, N_{\text{post}}\big)
\end{flalign} 
where $\eta_{\text{avg}}$ is the average transmittance and $N_{\text{post}}$ is the number of post-selected signals, obtained by applying threshold transmittance \cite{bib:21} filtering to the total number of pulses ($N$): 
\begin{equation} \eta_{\text{avg}} = \displaystyle \dfrac{\int_{\eta_{_\text{t}}}^{1} d\eta \; \eta \; p_{_{\eta_o, \sigma}}(\eta)}{\int_{\eta_{_\text{t}}}^{1} d\eta \; p_{_{\eta_o, \sigma}}(\eta)} \ ,  \ \       
    N_{\text{post}} = N \int_{  \eta_{_\text{t}}}^1 d\eta \; p_{_{\eta_o, \sigma}}(\eta)  \end{equation}
After postselection, a fraction of $ R_{\text{fin}}\big(\eta_{\text{avg}}\big)$ is left, giving a secure key rate (SKR) of \cite{bib:17}: 
\begin{equation} 
R_{\text{sec}} = R_{\text{fin}} \times \int_{  \eta_{_\text{t}}}^1 d\eta \; p_{_{\eta_o, \sigma}}(\eta)
\end{equation}
Here, the threshold transmittance is chosen such that the key rate is zero in the region where $\eta < \eta_{_\text{t}} $.  

The secure key length, $\ell$ produced from sending $N$ pulses is given as \cite{bib:19}: 
\begin{flalign}
\ell &= s_{_{X,0}} + s_{_{X,1}}  \\ & - s_{_{X,1}} h(\phi_{_{X}}) - n_{_X} f_{_{\text{EC}}} h(e_{\text{obs}}) - 6\log_2\frac{21}{\varepsilon_{{\text{sec}}}} - \log_2 \frac{2}{\varepsilon_{{\text{cor}}}} \nonumber
\end{flalign}
The first two terms represent the contributions to the key length from zero-photon and single-photon pulses, respectively. A fraction of the single-photon contribution $(s_{_{X,1}})$ is sacrificed for privacy amplification, as denoted by the \textit{third term,}. Here, $\phi_X$ represents the upper bound of the phase error rate in the rectilinear ($\boxplus$) basis and serves as the argument for the binary/Shannon entropy, $h(\cdot)$.

The \textit{fourth term} gives the expense for the error correction algorithm. In the $\boxplus$ basis, $n_X$ is the randomly chosen post-processing block size (for all $\mu_i$'s ) and $m_X$ is the number of incorrect bits. $f_{EC}$ denotes the error correction efficiency. In this case, the argument of the binary entropy function is the quantum bit error rate ($e_{obs} = \frac{m_X}{n_X}$). Finally, the last two terms account for relatively small errors, such as: probability of non-identical keys slipping through the error verification steps  and security leakage in the generated bit. We note that the entire diagonal basis ($\boxtimes$) is sacrificed between Alice and Bob to bound statistics in the $\boxplus$ basis.
 
 The final SKR is given by $R_{\text{fin}} = \frac{\ell}{N}$. This presents an optimization problem, as a lower threshold keeps more signal overall but also keeps signal with lower transmittance, which is more likely to be from noise in the detector. Meanwhile, a higher threshold reduces the amount of errors at the cost of sacrificing a larger portion of the key.

\begin{figure}[ht!]
    \centering
    \includegraphics[width=0.48\textwidth]{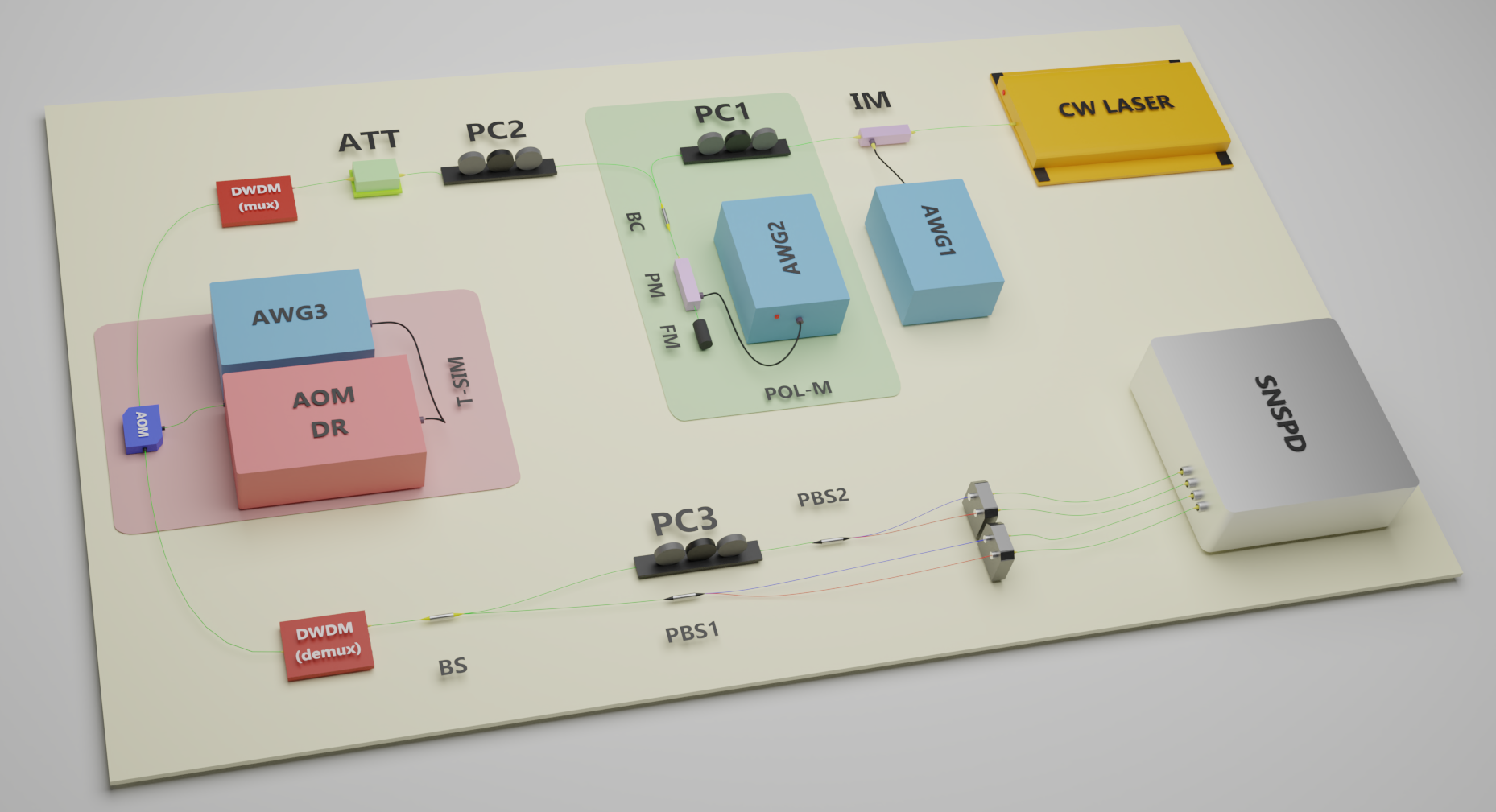}
    \caption{ Experimental setup of BB84 with decoy states. Alice prepares laser pulses using a continuous wave (CW LASER) source and intensity modulator (IM), driven by an arbitrary waveform generator (AWG1). Polarization encoding is done at polarization modulation (POL-M) stage. Optical fiber carrying the signal from one dense wavelength-division multiplexing (DWDM, mux) to other (DWDM, demux) is the quantum channel. The turbulence simulation stage (T-SIM) is to simulate atmospheric effect.   }
    \label{fig:example}
\end{figure}

\section{Experimental setup}

Our experimental setup is shown in Fig.\ \ref{fig:example}.

For weak coherent source, we used a low power (2 mW) continuous wave (CW) laser of central wavelength 1550.5 nm with high accuracy ($ \pm 2 \! \times \! 10^{-5}$ nm). We carved the continuous wave into pulses using a $\text{LiNbO}_3$ based intensity modulator. An arbitrary waveform generator (AWG1) was used to produce signal and decoy states. We used a null point modulator bias controller device to ensure a stable operation state by applying compensation bias voltage. 

We encoded each pulse with a polarization in either the $\boxplus$ or $\boxtimes$ basis (H, D $\equiv 0$; V, A $\equiv$ 1). Polarization modulation was achieved through the use of a beam circulator (BC), a phase modulator (PM) and a Faraday mirror (FM). The sequence of pulses traveled through a beam circulator and then a phase modulator which modified the phase using short voltage pulses with the help of an arbitrary waveform generator (AWG2). To address any undesired changes in the polarization states due to frequency or polarization mode dispersion, a FM was employed to compensate for any induced phase shift $\phi_{e}$.

After optical attenuation through a combination of a digital and analog attenuators, the pulse train was multiplexed through a dense wavelength division multiplexing (DWDM) device. Upon being demultiplexed with a corresponding DWDM at Bob's side, random basis selection ( $\boxplus$ or $\boxtimes$ ) was performed with a 50:50 beam splitter (BS) and polarization measurements ( $\leftrightarrow$, $\updownarrow$  \; or \; $\diagonal$, $\antidiagonal$ ) were performed with polarization beam splitters (PBS). Finally, a superconducting nanowire single photon detector (ID281 SNSPD) sent the detected signal to a time to digital converter (TDC, ID801) which was synced with AWG1, AWG2 and AWG3. 

Alice prepares her states by optimizing the decoy state parameters $\{ q_x, \mu_1, \mu_2, p_{\mu_1}, p_{\mu_2}\}$ \cite{bib:19} associated with the desired turbulence parameters $\eta_o$ and $\sigma$. The AWG1 uses $\mu_1$ and $\mu_2$ to implement the average intensity of the signal and weak decoy states on the pulses, respectively. The occurrence rates of these two states are controlled by $p_{\mu_1}$ and $p_{\mu_2}$ in order. The vacuum state intensity, $\mu_3$ is set to zero and the probability of occurrence, $p_{\mu_3}$ can be found from the relation: $p_{\mu_1} + p_{\mu_2} + p_{\mu_3} = 1$. The AWG2 controls the probability of Alice choosing either $\boxplus$ or $\boxtimes$ basis, utilizing the parameters $q_x$ and $q_z$. Note that $q_x + q_z =1$. 

Channel parameters $\eta_o$ and $\sigma$ are used to create an arb file for AWG3 to implement log-normally distributed transmittance for the quantum channel using AOM. 

The optimized parameter values for $37$ dB and $40$ dB cases are listed in Table \ref{tab:1}.

\begin{table}[H]
\caption{\raggedright \footnotesize Optimized parameter values obtained for turbulence parameter set $\{\eta_o, \sigma\}$, using the method in \cite{bib:19}. }
\label{tab:1}
    \begin{tabular}{c@{\hspace{8mm}} c@{\hspace{8mm}} c@{\hspace{8mm}} c@{\hspace{8mm}} c@{\hspace{7mm}} c}
    \hline \hline \addlinespace[1ex]
    Turbulence & $q_x$ & $\mu_1$ & $\mu_2$ & $p_{\mu_1}$ & $p_{\mu_2}$   \\ \addlinespace[1ex]
    \hline \addlinespace[1ex]
    37 dB   & 0.795 & 0.678 & 0.293 & 0.361 & 0.429 \\ 
    40 dB   & 0.677 & 0.701 & 0.281 & 0.246 & 0.490  \\ \addlinespace[1ex]  
    \hline \hline \addlinespace[1ex]      
    \end{tabular}
    \label{tab:my_label}
\end{table}

\section{Analysis}
Tables \ref{tab:2} and \ref{tab:3} show the critical differences between the old single-photon avalanche detectors (SPAD) used in \cite{bib:12}, and the new SNSPD detectors. The dark count rate of the new detectors is almost 2 orders of magnitude lower than that of the old detectors. The SNSPD has a dead time ranging from $70-80$ ns, whereas our old SPAD detector's dead time is $9.1$ $\mu$s. Compared to the old detectors, the after-pulse recovery time of the SNSPD detectors is significantly smaller. In our experiment, we used a 10 MHz signal, which results in a repetition rate of 100 ns. Consequently, there is no waiting time between two consecutive pulses, whereas the SPAD detectors had a significant recovery period between detections, resulting in lost pulses. 

\begin{figure}[ht!]
    \centering
    \includegraphics[width=0.5\textwidth]{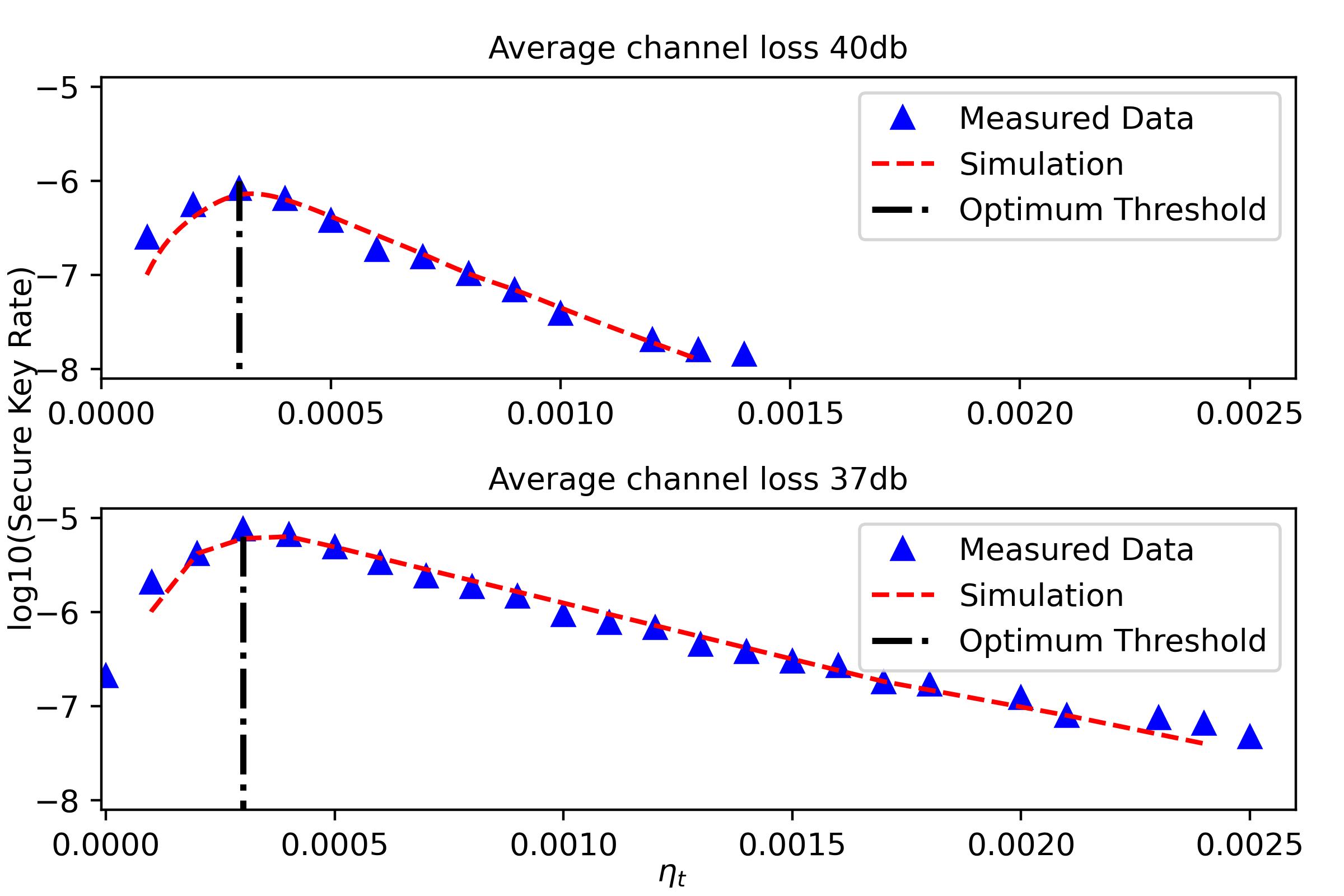}
    \caption{ Finding optimized key rate using ARTS \cite{bib:24} method. The variation in logarithm of the secure key rate for increasing applied transmittance cutoff: For 40 dB mean channel loss (top) and For 37 dB mean channel loss (bottom) }
    \label{fig:2}
\end{figure}

After preparing the setup, we collected $3\times10^{10}$ bits of data (processing time is approximately 50 minutes at the mentioned rate) for analysis. To find the optimized threshold region, we studied the distilled key rate as a function of threshold transmittance $\eta_{\text{t}}$ for both cases \cite{bib:24}. The result is shown in Fig.\ \ref{fig:2}.   

\begin{table}[ht]
\caption{\footnotesize Background-noise parameters comparison between old and new sets of detectors. Background click probability $P_{bg}(\eta) = Y_o + b\eta$. The SNSPD (new) is completely independent on transmittance. }
\label{tab:2}
\begin{tabular}{c@{\hspace{9mm}}  c@{\hspace{9mm}}  c@{\hspace{6mm}}  c@{\hspace{4mm}}  c@{\hspace{3mm}}}
\hline \hline \addlinespace[1.5ex]\; 
\multirow{2}{*}{Detector} & $Y_o^{\text{old}} \;\; $  & $Y_o^{\text{new}} \;\; $ & $ b^{\text{old}} \;\; $ & $b^{\text{new}} $ \\  \addlinespace[0ex]
 & \tiny $( \times 10^{-7})$ & \tiny $ (\times 10^{{-7}})$  & \tiny $(\times 10^{-4})$ & \tiny $(\times 10^{-4})$ \\ \addlinespace[1ex]
\hline \addlinespace[1ex]
$\leftrightarrow$   & $76 \pm 6$ & $7.1 \pm 0.6$ & $2.6 \pm 0.4$ & \\
$\updownarrow$     & $310 \pm 20$ & $6.7 \pm 0.6$ & $1.8 \pm 0.4$ & \multirow{2}{*}{0} \\
$\diagonal$     & $ 670 \pm 30$ & $6.2 \pm 0.6$ & $2.7 \pm 0.4$ & \\
$\antidiagonal$     & $670 \pm 30$ & $6.1 \pm 0.6$ & $1.8 \pm 0.4$ & \\ \addlinespace[1ex]
\hline \hline 
\end{tabular}
\end{table}

\begin{table}[ht]
\caption{\raggedright \footnotesize Comparison between old and new device parameters}
\label{tab:3}
\begin{tabular}{l@{\hspace{8mm}}  r@{\hspace{8mm}}  r}
\hline \hline \addlinespace[1ex]\; 
Experimental Parameters & Old Setup & New Setup \\\addlinespace[1ex] \hline \addlinespace[1ex]

Bob's optical efficiency & $0.42 \pm 0.02$ & $0.42 \pm 0.02$ \\ 
Optical misalignment & $0.003 \pm 0.002$ & $0.001 \pm 0.0004$ \\ 
Quantum efficiency & $0.1 \pm 0.05$ & $.8 \pm 0.05$ \\  
Dead time & $  \sim 9000 $ ns & $ \leq 80$ ns  \\ 
Time jitter & $\leq 200$ ps & $\leq 50$ ps \\ \addlinespace[1.5ex]

\hline \hline \\ 
\end{tabular}
\end{table}

In both situations, we found the optimized threshold lying between the range $2.5 \times10^{-4}$ to $3.5\times10^{-4}$. We chose $\eta_{\text{tc}}= 3\times10^{-4}$ as the optimum threshold for the next step. 

Here we calculated the the SKR as a function of mean channel loss in three different ways. We simulated $R_{\text{sec}} $ vs $\eta$ for both zero cutoff and for optimal cutoff conditions. The simulation graph shows substantial deviation ($\sim 1.85$ dB) for high-loss condition. Our measurement of key rate with considering $\eta_{\text{tc}}$ is very close to the simulation curve in both cases. The possible reasons for the deviation is optical misalignment and fluctuation in the average signal and decoy photon number.

\begin{figure}[H]
    \centering
    \includegraphics[width=0.95\columnwidth]{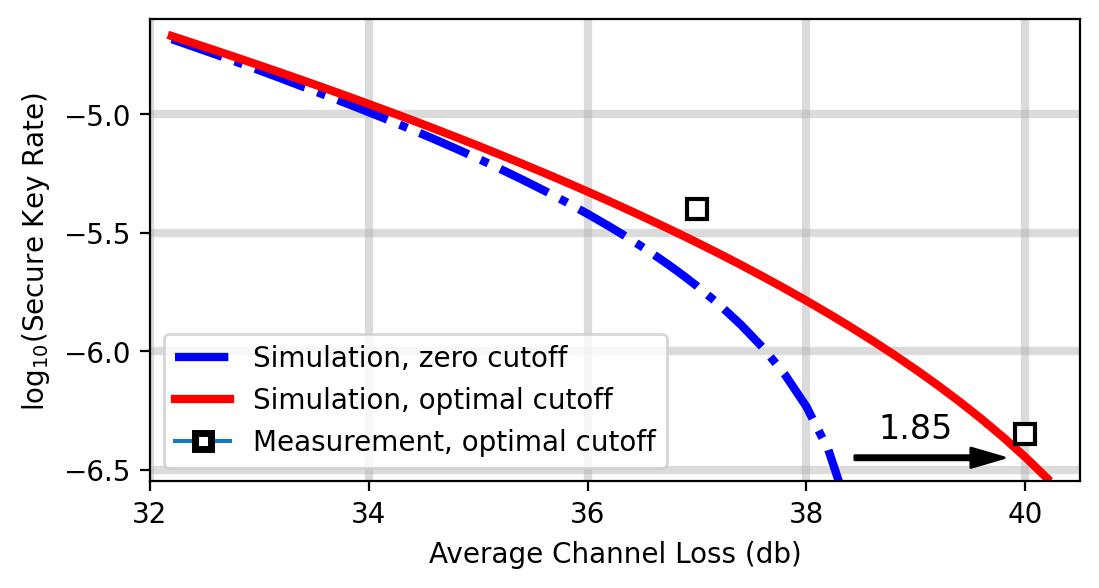}
    \caption{ Simulation and Measurements using P-RTS method: cutoff set at $\eta_{\text{ot}}$ = $3\times 10^{-4}$, taken from Fig.\ \ref{fig:2}. }
    \label{fig:3}
\end{figure}

By comparing the simulation and experimental graphs in Figs. \ref{fig:2} and \ref{fig:3}, we conclude:

\begin{enumerate}[(i)]
    \item Using threshold transmittance $\eta_{\text{t}}$ to discard bits at lower transmittance, we experimentally demonstrated the increment of the key rate beyond $40$ dB loss channel as proposed in \cite{bib:21} and demonstrated by our group up to 19 dB \cite{bib:12}.
    \item The optimum threshold region for maximum key rate is found to be similar for both the 40 dB and 37 dB channels, ranging from $2.5\times10^{-4}$ to $3.5\times10^{-4}$, and is independent of the channel parameters. This finding supports the P-RTS theory \cite{bib:17}. Our threshold region is significantly lower (by 2 orders of magnitude) than the previous findings reported in \cite{bib:12}, as expected since the new detectors' background click probability $P_{\text{bg}}$ is not dependent on transmittance. Moreover, the only contributing parameter to $P_{\text{bg}}$ is the background noise $Y_o$, which is also significantly smaller (by up to 2 orders of magnitude) (see Table \ref{tab:1}). The P-RTS paper predicts the dependability of $\eta_{\text{t}}$ only on device parameters, which provides additional evidence in favor of the theory. Interestingly, in our previous work, for the highest tolerated loss at 19 dB, the optimal threshold was found to be sensitive to channel statistics and experienced ~40\% decrease in SKR with a transmittance threshold optimized for 17 dB.
\end{enumerate}

\vspace{-2mm}

\section{Conclusion}
We experimentally simulated a finite-key decoy state BB84 in turbulent, high loss environments, of up 40 dB, which extended our previous work beyond 19 dB. This was achievable by replacing SPAD detectors with SNSPD detectors with higher efficiency, lower dark count rates and no after pulsing. This work is especially relevant for free space systems at the ground level, for example at metropolitan scales \cite{bib:23,bib:23a}.
 
We found that at the loss limits of our system, the optimal data rejection threshold was not sensitive to channel statistics, in accordance with the P-RTS theory in \cite{bib:17}. To extend our work, the effect of varying levels of turbulence can be analyzed in the context of P-RTS, which can also include testing higher turbulence levels ($\sigma^2 \! \gtrsim \! 1.2$) using the more appropriate gamma-gamma model, as suggested in \cite{bib:14}. Further studies can include experimental testing of other protocols such as measurement device independent quantum key distribution (MDI QKD) \cite{bib:22} over our simulated turbulent channel. An MDI QKD testbed could consider the effect of turbulence on channel asymmetry and its relation to the Hong-Ou-Mandel interferometer visibility at the detection apparatus, as has been investigated theoretically in \cite{bib:17}.

\section*{Acknowledgments}

We wish to thank B. Qi, E. Moschandreou, and B. Rollick for comments and discussion. 


\end{document}